\documentclass[12pt,english]{iopart}
\usepackage[english]{babel}
\usepackage{iopams} 
\usepackage{graphicx}

\def\Journal#1#2#3#4#5{{#1} #5 {\it #2} \textbf{#3} #4}

\newcommand{\K}{K_0}
\newcommand{\SI}{S_0}

\newcommand{\EE}{\mathbb{E}}
\newcommand{\EA}{\mathbb{E}_a}
\newcommand{\EB}{\mathbb{E}_b}
\newcommand{\EAB}{\mathbb{E}_{\ab}}
\newcommand{\PP}{\mathbb{P}}
\newcommand{\KK}{\mathbb{K}}
\newcommand{\uu}{\mathbf 1}
\newcommand{\FF}{\mathcal{F}\/}
\newcommand{\GG}{\mathcal{G}\/}
\newcommand{\ab}{a,b}
\newcommand{\tpm}{t^{\pm}}

\begin{document}

\title{Volatility and dividend risk in perpetual American options}
\author{Miquel Montero}
\address{Departament de F\'{\i}sica Fonamental, Universitat de Barcelona,\\  Diagonal 647, E-08028 Barcelona, Spain. }
\ead{miquel.montero@ub.edu}
\date{\today}

\begin{abstract}
American options are financial instruments that can be exercised at any time before expiration. In this paper we study the problem of pricing this kind of derivatives within a framework in which some of the properties ---volatility and dividend policy--- of the underlaying stock can change at a random instant of time, but in such a way that we can forecast their final values. Under this assumption we can model actual market conditions because some of the most relevant facts that may potentially affect a firm will entail sharp predictable effects. We will analyse the consequences of this potential risk on perpetual American derivatives, a topic connected with a wide class of recurrent problems in physics: holders of American options must look for the fair price and the optimal exercise strategy at once, a typical question of free absorbing boundaries. We present explicit solutions to the most common contract specifications and derive analytical expressions concerning the mean and higher moments of the exercise time.

\end{abstract}


\section{Introduction}
\label{S1}
Pricing financial derivatives is a main subject in mathematical finance with clear implications in physics. In 1900, five years before Einstein's classic paper, Bachelier~\cite{C64} proposed the {\it arithmetic\/} Brownian motion for the dynamical evolution of stock prices with the aim of obtaining a formula for option valuation. Samuelson~\cite{PAS65} noticed the structural failure of Bachelier's market model: it allowed negative values for the stock price, what led to undesired consequences in option prices. For correcting these unwanted features he introduced the {\it geometric\/} Brownian motion. Within his log-normal model, Samuelson obtained the fair price for perpetual options, although he was unable to find a general solution for expiring contracts. The answer to this question must wait until the publication of the works of Black and Scholes~\cite{BS73}, and Merton~\cite{M73}. The celebrated Black-Scholes formula has been broadly used by practitioners since then, mainly due to its unambiguous interpretation and mathematical simplicity. 

It is well established, however, that this model fails to fit some features of actual derivatives. In particular, there is solid evidence pointing to the necessity of relaxing the assumption, present in the Black-Scholes model, that a constant volatility parameter drives the stock price. Many models have been developed with the purpose of avoiding this restrictive condition: in Merton~\cite{M76} volatility was a deterministic function of time, in Cox and Ross~\cite{CR76} was stock-dependent, Hull and White~\cite{HW87} proposed a model where the squared volatility also follows a log-normal diffusion equation, Wiggins~\cite{W87} considered underlying and volatility as a two-dimensional system of correlated log-normal random processes, in Scott~\cite{S87}, and also in Stein and Stein~\cite{SS91}, the instantaneous volatility follows a mean-reverting arithmetic Ornstein-Uhlenbeck process, and Heston~\cite{H93} introduced correlation in the preceding model, just to name a few.

Another standard limitation of the Black-Scholes formula is that it is restricted to European derivatives: the option can be exercised at maturity only. However, most of the exchange-traded options are American: they can be exercised anytime during the life of the contract. Once again there is a clear connection with typical problems in physics: the earliest analysis of the issue of pricing American derivatives was formulated by McKean~\cite{McK65} as a free boundary problem for the heat equation. Kim~\cite{IJK90} provided an integral representation of the option price but, unfortunately, we have nowadays no explicit expression for the American counterpart of the Black-Scholes formula. The kernel of the problem is in that, in general, the optimal exercise boundary is implicitly defined by the integral equation that determines the price of the option. Only under certain circumstances closed expressions for American option prices do exist: e.g. in the case in which the properties of the derivative lead to a constant early exercise price, as in Rubinstein and Reiner~\cite{RR91}, or when the option is perpetual, as in Kim~\cite{IJK90}, and also in Elliott and Chan~\cite{EC04}, where the stock is driven by fractional Brownian motion. In the most general scenario analytical or numerical approximate methods must be used instead ---see, for instance, Barone-Adesi and Whaley~\cite{BAW87}, Broadie and Detemple~\cite{BD96}, Ju~\cite{NJ98}, Broadie {\it et al\/}~\cite{BDGT00}, and references therein as well.

In this article we will generalize a market model first introduced by Herzel~\cite{H98} as a simplified version of Naik's work. Naik~\cite{N93} developed a model in which the volatility can take only two known values, and the market switches back and forth between them in a random way. Herzel let the volatility jump at most once: a suitable way for encoding a market that may undergo a severe change in volatility only if some forthcoming event takes place. Herzel formally solved the problem of pricing European options if the market price of volatility risk was constant. The problem was revisited in~\cite{MM04}, where different risk premiums were considered and some explicit solutions were found. Here we will tackle the problem of pricing perpetual American options within a framework where volatility but also dividend rate may perform a single transition at some unknown instant in the future.

The paper is structured as follows: In \Sref{S2} we present the market model and its general properties. In \Sref{S3} we introduce the concept of financial derivative, and explore the links between finance and physics in the context of American derivatives. \Sref{S4} is devoted to the subject of pricing perpetual American options: we stress the financial interest of these ideal derivatives and emphasize the potentials of the analytical expressions found. In \Sref{S5} we show some illustrative examples and discuss their implications. Conclusions are drawn in \Sref{Last}, and the paper ends with two appendices. In~\ref{SA} we revisit the problem by following a different approach, and~\ref{SB} deals with the smooth pasting condition for vanilla options.  

\section{The market model}
\label{S2}
Let us begin with the general description of our set-up. We will consider a financial market where the non-deterministic stock $S$ is traded. The price evolution of this stock, assuming that $S=S(t_0)$ at $t=t_0$, is the following:   
\begin{equation*}
S(t)=S(t_0)+ \int_{t_0}^{t}\mu(t')S(t')  dt' +\int_{t_0}^{t}\sigma(t') S(t') dW(t'-t_0),\label{dS}
\end{equation*}
where $W(t)$ is a Wiener process, a one dimensional Brownian motion with zero mean and variance equal to $t$. The drift, $\mu$, and the volatility, $\sigma$, are stochastic quantities whose initial values are $\mu_a$ and $\sigma_a$. After $t_0$ they may simultaneously change to different fixed values, $\mu_b$ and $\sigma_b$, but such a transition can take place only once in a lifetime:  
\begin{equation*}
\mu(t)=\mu_a \uu_{t\leq \tau} +\mu_b \uu_{t>\tau},
\end{equation*}
\begin{equation*}
\sigma(t)=\sigma_a \uu_{t\leq \tau} +\sigma_b \uu_{t>\tau}.
\end{equation*}
Throughout the text $\uu_{\{\cdot\}}$ will denote the indicator function, which assigns the value 1 to a true statement, and the value 0 to a false statement. Note that if $\sigma_a \neq \sigma_b$, we will have a market model with stochastic volatility. The case in which $\mu_a \neq \mu_b$ has in principle a more flexible interpretation, although we will concentrate our attention in the existence of two different (continuous in time) dividend pay-off regimes:
\begin{equation*}
\delta(t)=\delta_a \uu_{t\leq \tau} +\delta_b \uu_{t>\tau}.
\end{equation*}
These magnitudes are stochastic because the instant $\tau>t_0$ in which the transition occurs is a random variable. We will assume that $\tau$ follows an exponential law:
\begin{equation}
\PP\{t_0<\tau\leq t|\FF(t)\}=1-e^{-\lambda (t-t_0)}. 
\label{PP}
\end{equation}
Note that~(\ref{PP}) identifies $\lambda$ as the inverse of the mean value of the transition waiting time interval $\tau-t_0$, $\EE[\tau-t_0|\FF(t_0)]=\lambda^{-1}$. In the previous expressions and hereafter $\FF(t)$ represents {\it all\/} the available information up to time $t$. 

The asset behaviour may be easily visualized when we express it in terms of returns, $R(t;t_0)\equiv \log(S(t)/S(t_0))$, instead of spot prices. The return will follow a drifted Wiener process with parameters $(\mu_a,\sigma_a)$ up to time $\tau$. After that time the initial Brownian motion freezes and a second drifted Wiener process drives the subsequent evolution of the return:    
\begin{equation*}
R(t;t_0)=R_a(\min(t,\tau);t_0)+R_b(\max(t,\tau);\tau),
\end{equation*}
with $R_{\ab}(t;t')=\sigma_{\ab} W_{\ab}(t-t')+\tilde{\theta}_{\ab}(t-t')$, and $\tilde{\theta}_{\ab}=\mu_{\ab}-\sigma_{\ab}^2/2$. 

In the most common situation the stock $S$ and its derivatives, contracts whose price depends upon the value of this underlying, are the only securities affected by the actual value of $\tau$. When part of the risk is not directly traded in the market, the market may be incomplete: we will not be able to reproduce the behaviour of some assets by means of a {\it replicating\/} portfolio. In our case, if we want to hedge the market exposure of derivatives to volatility or dividend risk we must also include in the portfolio {\it secondary derivatives\/}, derivatives with the same underlying stock but different contract specifications. The immediate consequence of such constraint is that the risk premium coming from $\tau$ is arbitrary to a certain extent, because investors can evaluate it on the basis of their own perceptions. 
We will avoid entering into the discussion of the financial consequences of this arbitrariness now: we leave it to~\ref{SA}, where the hedging-portfolio approach is taken. 
The most relevant point to be noted here is that we can use~(\ref{PP}) in order to define a risk-neutral measure for $S$, 
\begin{eqnarray*}
\PP\{s<S(t)\leq s +ds|\FF(t_0)\}=\left\{
\frac{e^{-\lambda (t-t_0)}}{s \sqrt{2\pi \sigma^2_a (t-t_0)}} e^{-\frac{[\log(s/S(t_0))-\theta_a(t-t_0)]^2}{2 \sigma_a^2 (t-t_0)}} \right.\\
+\left. \int_0^{t-t_0} du  \frac{\lambda e^{-\lambda u}}{s \sqrt{2\pi \bar{\sigma}^2(u,t-t_0) (t-t_0)}} e^{-\frac{[\log(s/S(t_0))-\bar{\theta}(u,t-t_0)(t-t_0)]^2}{2  \bar{\sigma}^2(u,t-t_0)(t-t_0)}}\right\}ds,
\end{eqnarray*}
where we have introduced some quantities depending on the risk-free interest rate $r$, the volatilities $\sigma_{\ab}$, and the dividend pay-offs $\delta_{\ab}$:
\begin{eqnarray*}
\theta_{\ab}=r-\delta_{\ab}-\sigma_{\ab}^2/2, \\
\bar{\theta}(u,v)=\frac{\theta_{a} u+\theta_{b}(v-u)}{v} \mbox{ and} \\
\bar{\sigma^2}(u,v)=\frac{\sigma^2_{a} u+\sigma^2_{b}(v-u)}{v}.
\end{eqnarray*}
The use of this measure guarantees that
\begin{equation}
F(t)=e^{-\int_{t_0}^{t}(r-\delta)dt'} S(t)
\label{defF}
\end{equation}
fulfils 
\begin{equation}
\EE[F(t)|\FF(t_0)]=F(t_0)=S(t_0). 
\label{martF}
\end{equation}
Note that physical and risk-free measures coincide when $\tilde{\theta}_{\ab}=\theta_{\ab}$.
\section{American option as a first passage problem}
\label{S3}

Options are contracts between two parties, sold by one party to another, that give the buyer the right, but not the obligation, to buy (call) or sell (put) shares of the underlying stock at some prearranged price, the {\it strike\/} price $K$, within a certain period or on a specific date, the maturity or expiration time $T$. Sometimes $K$ is a parameter but in general, depending on the contract specifications, it will be a function involving some other constants: 
\begin{equation}
K^{\pm}=S(t)\mp X_0 \uu_{S(t) \gtreqless \K}.
\label{Kb}
\end{equation}
We will use the generic sign $\KK$ as a shorthand for all the contract parameters. 
As a consequence of their privileged position, option holders will only exercise their rights if they obtain a net benefit. In other words, we can see options as contingent claims with their present value determined by the discounted value of the expected profit under our risk-neutral measure: 
\begin{equation*}
P(t_0,S(t_0);\KK)=\EE [X(S(t^*);\KK) e^{-r(t^*-t_0)}|\FF(t_0)],
\end{equation*}
where $X(S(t);\KK)$ is the pay-off function and $t^*$ is the actual exercise time.

The notation we use in the definition of the pay-off function is not incidental. We will concentrate our attention on those contracts for which the pay-off is a function of the current value of the asset, like in the case of vanilla calls (+) and puts (-) where
\begin{equation}
X^{\pm}(S;K,T)=\max(\pm(S-K),0) \uu_{t \leq  T},
\label{payv}
\end{equation}
and $K$ is constant. Another typical pay-off is 
\begin{equation}
X^{\pm}(S;\K,T)=X_0 \uu_{S \gtreqless  \K} \uu_{t \leq  T}.
\label{payb}
\end{equation}
We can derive this pay-off from~(\ref{Kb}) and~(\ref{payv}): it corresponds to binary or digital options. For the sake of simplicity we will set $X_0=1$ hereafter. Vanilla and binary options will be the only instances we will study in practice, although some other contracts may fit our requirements as well. Note that this is not the case of any exotic derivative whose pay-off depends on the past path of the stock, like in Asian options, Lookback options or {\it knock-out\/} options.   

When the option can be exercised at the end of the contract lifetime $T$ only, the exercise time is deterministic, and the option is said to be European. If the option can be exercised at any time before expiration it is called American, and $t^*$ becomes an stochastic magnitude as well. Note that the contract is always worthless after maturity: the option buyer must decide under which conditions the option can be optimally exercised before this deadline. The decision will finally depend on the present value of stock price $S(t)$ and the time to expiration, $T-t$. The problem is thus in essence a typical problem of first-passage time: we must determine at what time the process $S(t)$ will touch the boundary $H(t)$. In financial language, $H(t)$ is named the optimal exercise boundary, the stock price above (below) which it is better to exercise the call (put) than to keep the option alive. 
We will define the exercise time $\tpm(t_0)$ as the first time the underlying crosses the threshold given that at present time, $t_0$, the spot price of the asset lies in the proper side of the boundary:  
\begin{equation}
\tpm(t_0)=\min\left\{t > t_0; S(t) \gtreqless H(t)|S(t_0) \lessgtr H(t_0)\right\}.
\label{tpm}
\end{equation} 

The optimal strategy for choosing the boundary function derives from the following constraint: the investor must settle $H(t)$ in such a way that the value of the alive option equals the pay-off of the contingent claim in the optimal exercise price,
\begin{equation}
P^{\pm}(t,S=H;\KK)=X^{\pm}(H;\KK).
\label{continuity}
\end{equation}
The condition must be fulfiled in a smooth way as well, the smooth pasting condition~\cite{PAS65,M73,SS05}:
\begin{equation}
\left. \frac{\partial P^{\pm}(t,S;\KK)}{\partial S}\right|_{S=H}=\left. \frac{\partial X^{\pm}(S;\KK)}{\partial S}\right|_{S=H},
\label{smooth}
\end{equation}
in the case in which the right-hand side of the previous expression does exist. That is, the option price must be continuous with continuous derivative in the asset price when it crosses the boundary.
In conclusion, the investor must compute function $H(t)$ at the same time he or she evaluates the option price.

\section{Perpetual American options}
\label{S4}

We analyse now the problem of valuating perpetual American options, i.e. when we have $T -t_0\rightarrow \infty$. 
It can be objected that perpetual American options have limited practical interest since, in general, actual derivatives expire. In this sense, one can argue that they represent the limiting value of a far from maturity contract, and therefore they may help in the pricing process if the theoretical price cannot be computed~\cite{BAW87}. On the other hand, the existence of a fixed expiration time is a feature which is not shared by systems coming from other branches of science. The results we introduce in this section may be thus of interest in different fields. We will stress this interpretation later. 

The major simplification that perpetual American options bring is that the value of the boundary must be piece-like constant, 
\begin{equation*}
H(t)=H_a \uu_{t\leq \tau} +H_b \uu_{t>\tau},
\end{equation*}
given that the problem is stationary. Then, since the process $S(t)$ is continuous, we will have:
\begin{eqnarray}
P^{\pm}(t_0,\SI;\KK)&=&
X^{\pm}(H_a;\KK) \EE [e^{-r(\tpm_0-t_0)}\uu_{\tpm_0\leq \tau}|\FF(t_0)]\nonumber \\&+&X^{\pm}(H_b;\KK) \EE [e^{-r(\tpm_0-t_0)}\uu_{\tpm_0 > \tau}|\FF(t_0)], \label{Ppm}
\end{eqnarray}
with $\tpm_0=\tpm(t_0)$ and $\SI=S(t_0)$. The problem is simpler in the case of binary options since there is no financial reason for holding the option alive once we are in the bonus region. This implies that $H_{\ab}=\K$ and therefore:
\begin{equation}
D^{\pm}(t_0,\SI;\K)=\EE[e^{-r(\tpm_0-t_0)}|\FF(t_0)]. \label{Dpm}
\end{equation}
Here it is interesting to note that the magnitude we must compute for obtaining the price of the option is nothing but a typical moment-generating function of the first-passage time interval $(\tpm_0-t_0)$:
\begin{equation}
\EE [(\tpm_0-t_0)^n|\FF(t_0)] =
(-1)^n \left. \frac{\partial^n}{\partial r^n} \EE [e^{-r(\tpm_0-t_0)}|\FF(t_0)] \right|_{r=0}, (n>0). \label{moments}
\end{equation}
In order to conserve this extra functionality in the output of the problem under analysis, we will keep $\theta_a$ and $\theta_b$ unexplicit as much as we can. In this way, we can turn risk-neutral results into physical results merely by replacing $\theta_{\ab}$ with $\tilde{\theta}_{\ab}$.

\subsection{The constant case}
Let us consider in the first place the case in which drift and volatility are constant. For $\bar{\theta}=\theta_{\ab}$ and $\bar{\sigma}=\sigma_{\ab}$ the probability density function (pdf) of $\tpm_0$ 
can be obtained by invoking the reflection principle of the Brownian motion:
\begin{eqnarray*}
\PP\{t<\tpm_0\leq t +dt|\FF(t_0)\}=\\\psi^{\pm}_{\ab}(t;t_0,\SI,H_{\ab})dt=
\frac{\pm x_{\ab}}{\sqrt{2\pi \sigma^2_{\ab} (t-t_0)^3}} e^{-\frac{(x_{\ab}-\theta_{\ab} (t-t_0))^2}{2 \sigma_{\ab}^2 (t-t_0)}}dt,
\end{eqnarray*}
where $x_{\ab}=\log(H_{\ab}/\SI)$. 
Under this assumption, the value of~(\ref{Ppm}) is well known~\cite{BAW87,W98}: 
\begin{eqnarray}
P^{\pm}(t_0,S_0,\KK) &=&X^{\pm}(H_{\ab};\KK)
\EAB[e^{-r(\tpm_0-t_0)}|\FF(t_0)]\nonumber \\&=&X^{\pm}(H_{\ab};\KK)\left(\frac{\SI}{H_{\ab}}\right)^{\beta^{\pm}_{\ab}}.
\label{Eab}
\end{eqnarray}
Here, with $\EAB[\cdot|\FF(t_0)]$ we mean that we are using $\psi^{\pm}_{\ab}(t;t_0,\SI,H_{\ab})$ in the computation of expected values. We have also introduced constants $\beta^{\pm}_{\ab}$, 
\begin{equation}
\beta^{\pm}_{\ab} = \frac{1}{\sigma^2_{\ab}}\left(-\theta_{\ab} \pm \sqrt{\theta^2_{\ab}+2 r \sigma^2_{\ab}}\right),
\label{betapm}
\end{equation}
which differentiate the overall properties of calls and puts since $\beta^{\pm}_{\ab}\gtrless 0$~\cite{BAW87}. Note that we must eventually recover~(\ref{Eab}) with $\beta^{\pm}_{a}$ ($\beta^{\pm}_b$) for the limiting case $\lambda \rightarrow 0$ ($\lambda \rightarrow \infty$). 
When the pay-off is~(\ref{payv}) we will have a perpetual American vanilla option: 
\begin{equation*}
V^{\pm}_b(t_0,\SI;K)=\pm(H_b - K) \left(\frac{\SI}{H_b}\right)^{\beta^{\pm}_b}.
\end{equation*}
The value of $H_b$ shall be obtained by demanding smoothness of the solution, recall~(\ref{smooth}), what implies here that
\begin{equation*}
{\left.\frac{\partial V^{\pm}_b(t,S;K)}{\partial S} \right|}_{S=H_b}=\pm 1,
\end{equation*}
and therefore~\cite{BAW87}:
\begin{equation}
H_b = \frac{\beta^{\pm}_b}{\beta^{\pm}_b - 1} K. \label{Hb}
\end{equation}
For put options we have guaranteed $0<H_b<K$ because $\beta^{-}_b<0$. For call options it can be proved that $\beta^{+}_b \geq 1$, an therefore we will have $H_b>K$. In fact, for $\beta^{+}_b = 1$, or what is the same, for $\theta_b=r-\sigma^2_b/2$, the value of $H_b$ diverges: there is no optimal boundary, the option is never exercised and, as a consequence, it must quote as its underlying stock, $V^{+}_b(t,S(t))=S(t)$. Thus, if the volatility is constant and the stock pays no dividend, only American puts are meaningful. For them we will have $\beta^{-}_b=-2r/\sigma^2_b$, and therefore
\begin{equation*}
V^{-}_b(t_0,\SI;K)=\frac{\sigma^2_b\SI}{2 r} \left(\frac{H_b}{\SI}\right)^{1+\frac{2r}{\sigma^2_b}},
\end{equation*}
with
\begin{equation*}
H_b = \frac{K}{1+\frac{\sigma^2_b}{2 r}}.
\end{equation*}

\subsection{The general case}
We will now return to the most general case with the previous results in mind. Note that~(\ref{Ppm}) splits into two terms, the first one will count the realizations of the process that reach exercise price {\it before\/} the change in the dynamics, and the second one collects those for which the optimal boundary is hit {\it after\/} it.
The first expectation can be reduced to: 
\begin{equation*}
\EE [e^{-r(\tpm_0-t_0)}\uu_{\tpm_0 \leq \tau}|\FF(t_0)]=\EA[e^{-(r+\lambda)(\tpm_0-t_0)}|\FF(t_0)],
\end{equation*}
because
\begin{equation*}
\EE [\uu_{\tpm_0 \leq \tau}|\FF(\tpm_0)]=1-\PP\{\tau < \tpm_0|\FF(\tpm_0)\}=e^{-\lambda(\tpm_0-t_0)},
\end{equation*}
and the pdf of $\tpm_0$ 
is $\psi^{\pm}_a(t;t_0,\SI,H_a)$, provided that $\tpm_0 \leq \tau $. 
Therefore, we can adapt the result in~(\ref{Eab}) in order to obtain
\begin{equation*}
\EA[e^{-(r+\lambda)(\tpm_0-t_0)}|\FF(t_0)]=\left(\frac{\SI}{H_a}\right)^{\gamma^{\pm}_{a}},
\end{equation*}
where we have introduced $\gamma^{\pm}_{a}$ which depends on $\lambda$:
\begin{equation}
\gamma^{\pm}_{a}=\frac{1}{\sigma^2_a}\left(-\theta_a \pm \sqrt{\theta^2_a+2 (r+\lambda) \sigma^2_a}\right).
\label{gamma}
\end{equation}

The computation of the second term in~(\ref{Ppm}) is cumbersome. We will address a simpler problem in the first place, the calculation of $\EE [e^{-r(\tpm_0-t_0)}\uu_{\tpm_0 > \tau}|\FF(\tau)]$, and after that we will recover the true expression thanks to the following identity:
\begin{equation}
\EE [e^{-r(\tpm_0-t_0)}\uu_{\tpm_0> \tau}|\FF(t_0)]= 
\EE \left[\left.\EE [e^{-r(\tpm_0-t_0)}\uu_{\tpm_0> \tau}|\FF(\tau)]\right|\FF(t_0)\right].
\label{EEEE}
\end{equation}
In order to compute $\EE [e^{-r(\tpm_0-t_0)}\uu_{\tpm_0> \tau}|\FF(\tau)]$ we must take into account that the indicator selects those trajectories for which either
$\widehat{S}^{+}(t)=\max_{t_0\leq t'\leq t} S(t')$, the maximum value of the process, or $\widehat{S}^{-}(t)=\min_{t_0\leq t'\leq t} S(t')$, its minimum value, has not met the optimal boundary at or before $\tau$: $\widehat{S}^{\pm}(\tau)\lessgtr H_a $. In this case the problem of finding the passage time renews:
\begin{eqnarray}
\EE [e^{-r(\tpm_0-t_0)}\uu_{\tpm_0> \tau}|\FF(\tau)]= 
e^{-r(\tau-t_0)}\EE [e^{-r(\tpm(\tau)-\tau)}|\FF(\tau)]\uu_{\widehat{S}^{\pm}(\tau)\lessgtr H_a },
\label{Ftau}
\end{eqnarray}
where the definition of $\tpm(\tau)$ follows from~(\ref{tpm}).

The expected value in the right-hand side of~(\ref{Ftau}) is straightforward ---cf~(\ref{Eab}):
\begin{eqnarray*}
\EE[e^{-r(\tpm(\tau)-\tau)}|\FF(\tau)]=\EB[e^{-r(\tpm(\tau)-\tau)}|\FF(\tau)]=
\left(\frac{S(\tau)}{H_b}\right)^{\beta^{\pm}_b}.
\end{eqnarray*}
Therefore, we have to obtain the probability density function of $S(\tau)$, $S(\tau)=\SI e^{R(\tau;t_0)}$, under the  restriction $\widehat{S}^{\pm}(\tau)\lessgtr H_a$, in order to complete the computation of~(\ref{EEEE}). Note that, under the risk-neutral measure, we can express the return up to time $\tau$ as $R(\tau;t_0)=\pm \sigma_a \widetilde{W}^{\pm}_a(\tau;t_0)$, with
\begin{equation*}
\widetilde{W}^{\pm}_a(t;t_0)=\pm \left[W_a(t-t_0)+\frac{\theta_a}{\sigma_a}(t-t_0) \right],
\end{equation*}
a Wiener process with a drift. In these terms both conditions
$\widehat{S}^{\pm}(\tau)\lessgtr H_a$ are equivalent to demand that the maximum of the Wiener process $\widetilde{M}^{\pm}_a(\tau;t_0)$ fulfils
\begin{equation*}
\widetilde{M}^{\pm}_a(\tau;t_0)=\max_{t_0\leq t'\leq \tau} \widetilde{W}^{\pm}_a(t';t_0) < \pm \frac{x_a}{\sigma_a}.
\end{equation*}
Thus, we have to compute the joint probability density function $\phi^{\pm}_a(m,w,t,t')$ of the two random processes,
\begin{eqnarray*}
\PP\{ m<\widetilde{M}^{\pm}_a(t;t')\leq m+dm, w<\widetilde{W}^{\pm}_a(t;t')\leq w+dw\}\\
=\phi^{\pm}_a(m,w,t,t')dm dw,
\end{eqnarray*}
which reads: 
\begin{eqnarray*}
\phi^{\pm}_a(m,w,t,t')=\nonumber \\
\frac{2(2m-w)}{\sqrt{2 \pi (t-t')^3}}\exp\left\{-\frac{(2m-w)^2}{2(t-t')} \pm\frac{\theta_a}{\sigma_a}w -\frac{\theta^2_a}{2 \sigma^2_a}(t-t')\right\}.
\end{eqnarray*}
Therefore
\begin{eqnarray*}
\EE [e^{-r(\tpm-t_0)}\uu_{\tpm_0> \tau}|\FF(t_0)]=  
\frac{\lambda}{\lambda+\ell^{\pm}} \left[\left(\frac{\SI}{H_b}\right)^{\beta^{\pm}_b} -\left(\frac{H_a}{H_b}\right)^{\beta^{\pm}_b} \left(\frac{\SI}{H_a}\right)^{\gamma^{\pm}_{a}} \right],
\end{eqnarray*}
where we have introduced another constant depending on the parameters of the problem:
\begin{eqnarray}
\ell^{\pm} &=& \left(\beta^{\pm}_a-\beta^{\pm}_b\right)\left(\frac{r}{\beta^{\pm}_a}+\frac{1}{2}\sigma^2_a\beta^{\pm}_b\right), 
\label{ell}
\end{eqnarray}
which cancels if no change in the market is induced by the point process $\tau$. 
When $H_{\ab}=\K$, as in the case of binary options, the sum of the two expectations lead to
\begin{equation}
\EE [e^{-r(\tpm_0-t_0)}|\FF(t_0)]= 
\frac{1}{\lambda+\ell^{\pm}} \left[\lambda\left(\frac{\SI}{\K}\right)^{\beta^{\pm}_b}  
+\ell^{\pm}\left(\frac{\SI}{\K}\right)^{\gamma^{\pm}_{a}} \right].\label{moment-generating}
\end{equation}
It is easy to check that in the two limiting cases, $\lambda \rightarrow 0$ and $\lambda \rightarrow \infty$, we recover the right expressions. In the case in which the pay-off function is~(\ref{payv}) we will have
\begin{eqnarray}
V^{\pm}(t_0,\SI;K) =
\pm(H_a - K)\left(\frac{\SI}{H_a}\right)^{\gamma^{\pm}_{a}} \nonumber \\  \pm(H_b - K)\frac{\lambda}{\lambda+\ell^{\pm}} \left[\left(\frac{\SI}{H_b}\right)^{\beta^{\pm}_b} -\left(\frac{H_a}{H_b}\right)^{\beta^{\pm}_b} \left(\frac{\SI}{H_a}\right)^{\gamma^{\pm}_{a}} \right].
\label{Vpm}
\end{eqnarray}
The value of $H_a$ must fulfil the transcendental equation that follows from the smoothness condition~(\ref{smooth}),
\begin{equation}
H_a (\gamma^{\pm}_a-1)-(H_b-K)  \frac{\lambda}{\theta_a+\sigma^2_a(\gamma^{\pm}_a+\beta^{\pm}_b)/2} \left(\frac{H_a}{H_b}\right)^{\beta^{\pm}_b}-\gamma^{\pm}_a K=0, \label{HH}
\end{equation}
where $H_b$ was introduced in~(\ref{Hb}).
We have seen above how an American call option quotes as its underlying if the volatility remains constant and the share pays no dividend. This conclusion is still valid in our framework, even though $\sigma_a\neq \sigma_b$. Let us consider in the first place $\delta_a\neq0$ and $\delta_b=0$. Here $\beta^{+}_b=1$, $H_b\rightarrow \infty$, and $\ell^{+}=\delta_a$. This reduces~(\ref{HH}) to an algebraic equation in $H_a$, which takes the simple form:
\begin{equation*}
H_a = \frac{\gamma^{+}_a(1+\lambda/\delta_a)}{\gamma^{+}_a-1} K > K.
\end{equation*}
Then 
\begin{equation*}
V^{+}(t_0,\SI;H_a)= \frac{\SI}{1+ \lambda/\delta_a} \left[\frac{1}{\gamma^{+}_a}\left(\frac{\SI}{H_a}\right)^{\gamma^{+}_{a}-1} +\frac{\lambda}{\delta_a}\right]<\SI,
\end{equation*}
for $\SI \leq H_a$. If we consider now the limit $\delta_a\rightarrow 0$ we can see that we recover $H_a\rightarrow \infty$ and the option is never exercised also in this case. The opposite situation, in which $\delta_a=0$ but $\delta_b\neq0$, is much more complex.
We have evidences pointing to the conclusion that $H_a\rightarrow \infty$ for any combination of the remaining parameters ---see~\ref{SB}. 
This implies that the stock must pay dividends after the change or in both periods in order to differ call options from their underlying with certainty. 

In the case of American puts we can still consider volatility and dividend risks separately: $H_b\neq 0$ and it is bounded since $\beta^{+}_b<0$, and $H_a=0$ does not solve equation~(\ref{HH}) for $\lambda \neq 0$. Moreover, when $\lambda =0$ the expression for $H_a$ trivially reduces to:
\begin{equation}
H_a = \frac{\beta^{\pm}_a}{\beta^{\pm}_a - 1} K.
\label{Ha}
\end{equation}
For vanilla put options, as well as for call options with $\beta^{+}_b\neq 1$, the most relevant feature of transcendental equation~(\ref{HH}) is the evident fact that it formally admits multiple solutions. A closer analysis shows that~(\ref{HH}) may have two real and positive solutions {\it at the most\/}. In~\ref{SB} we explain how in such a case we can always discard one them eventually. We also find the right expression in a practical situation involving put options. The most surprising fact is that there are market conditions for which equation~(\ref{HH}) has no real solution. Unfortunately we have found no financial reason that motivates this behaviour. In the next section we will consider a sample case that exhibits this handicap but shows no other evident singularity.      

Finally note that the above pricing expressions assume that $\gamma^{\pm}_a\neq \beta^{\pm}_b$. In order to obtain the right limiting values when a numerical concordance in the parameters lead to $\gamma^{\pm}_a = \beta^{\pm}_b$, we have to take into account that $\gamma^{\pm}_a \rightarrow \beta^{\pm}_b$ implies that $\ell^{\pm}\rightarrow -\lambda$ too.  In such a case one can easily check that, for instance, the solution in~(\ref{Vpm}) must be replaced by:  
\begin{eqnarray*}
V^{\pm}(t_0,\SI;H_a,K)=\\ \pm\left[(H_a - K) \left(\frac{\SI}{H_a}\right)^{\beta^{\pm}_b} -(H_b - K)\frac{\lambda}{\theta_a+\sigma^2_a \beta^{\pm}_b}\left(\frac{\SI}{H_b}\right)^{\beta^{\pm}_b} \log\left(\frac{\SI}{H_a}\right)\right].
\end{eqnarray*}

\section{Some examples and applications}
\label{S5}
In this section we present some practical examples that may illustrate the consequences of volatility and dividend risk in the properties of perpetual American options. We will consider, in the first place, the pay-off~(\ref{payv}) and analyse the outcomes related to the price.

\begin{figure}[tbh] 
{\hfil
\includegraphics[width=0.60\textwidth,keepaspectratio=true]{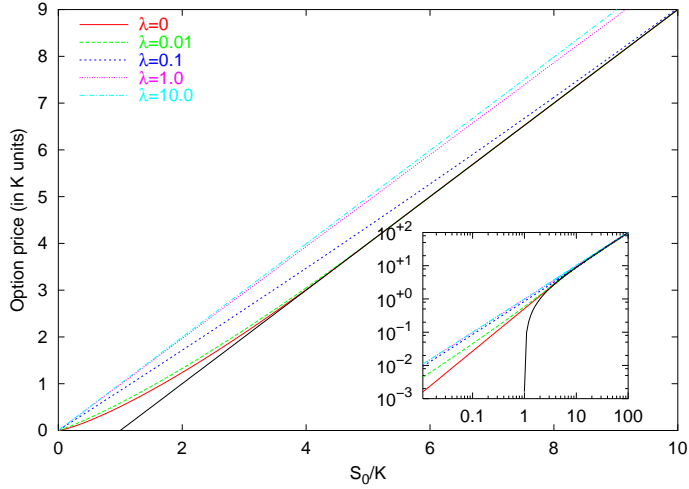} 
}
\caption{Option prices for a perpetual vanilla call under dividend risk. Here we consider the implications of a possible abrupt stoppage in the dividend payment. We represent the price of the option, in terms of the moneyness ($S_0/K$), for different values of $\lambda$. We have used typical market values for the parameters: $r=4\%$, $\delta_a=1.75\%$, $\delta_b=0\%$ and $\sigma_a=\sigma_b=25\%$. We plot as well the pay-off function in a solid (black) line.}
\label{figVCM}
\end{figure}

In figure~\ref{figVCM} we can see the effect of a change in the dividend policy on call prices. In particular we consider that the quoted firm may suddenly stop the dividend payment to the shareholders. The consequence is that the value of the option increases due to the possibility of a stoppage in the distribution of dividends. The optimal exercise price also increases with respect to the undisturbed one: note that $\lambda=0$ represents the limiting situation in which the change is impossible. When $\lambda$ grows the value of the option attains progressively the stock price, which correspond to the limiting case $\lambda\rightarrow \infty$, since here $\delta_b=0$. For finite values of $\lambda$, however, all prices must eventually converge to the pay-off function: we can check this property in the inset of figure~\ref{figVCM}.

\begin{figure}[tbh] 
{\hfil
\includegraphics[width=0.60\textwidth,keepaspectratio=true]{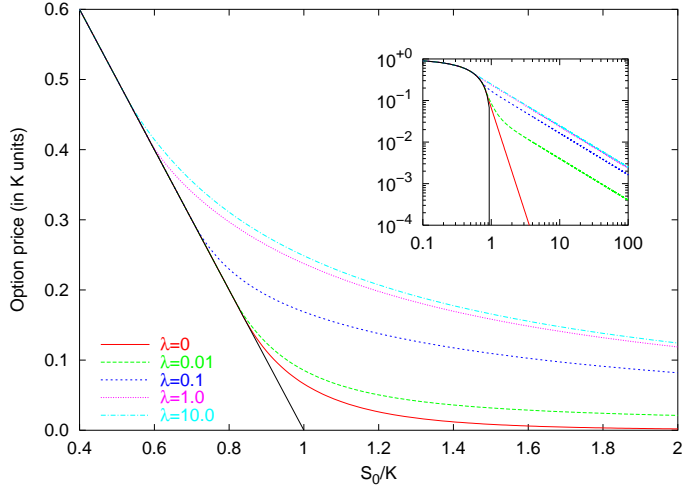} 
}
\caption{Option prices for a perpetual vanilla put under volatility risk.
We represent the price of the option, in terms of the moneyness ($S_0/K$), for different values of $\lambda$. We analyse here the consequences of a sudden increment in the volatility of the stock. We have used the following values for the parameters: $r=4\%$, $\delta_a=\delta_b=1.75\%$, $\sigma_a=10\%$ and $\sigma_b=25\%$.  The pay-off function is depicted in a solid (black) line.}
\label{figVPM}
\end{figure}

\begin{figure}[tbh] 
{\hfil
\includegraphics[width=0.60\textwidth,keepaspectratio=true]{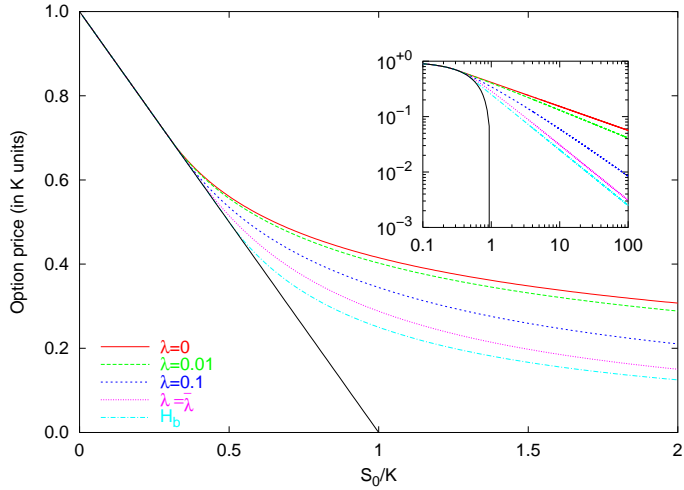} 
}
\caption{Option prices for a perpetual vanilla put under volatility risk.
We represent the price of the option, in terms of the moneyness ($S_0/K$), for different values of $\lambda$. We analyse here the consequences of a severe reduction in the volatility of the stock. We have used the following values for the parameters: $r=4\%$, $\delta_a=\delta_b=1.75\%$, $\sigma_a=40\%$ and $\sigma_b=25\%$.  The previous market conditions lead to the existence of threshold $\bar{\lambda}\approx 0.509$. We also plot the theoretical limiting case $\lambda \rightarrow \infty$ for which $H_a$ should equal $H_b$.}
\label{figVPMX}
\end{figure}

The following two examples deal with the effect of a change in the volatility on put prices. In figure~\ref{figVPM} we can see how the price of the put steadily increases with the likelihood of a sudden growth in the volatility level. The behaviour reverses in figure~\ref{figVPMX}, where a possible change of volatility would imply a reduction in the put value. Note that we are assuming in this figure a market set-up which lead to the presence of a maximum admissible value for $\lambda$,  $\bar{\lambda}\approx 0.509$, with no clear reason ---see~\ref{SB}. 

Another interesting property of option put prices (not related to $\bar{\lambda}$) is the following: If we observe the insets of figures~\ref{figVPM} and~\ref{figVPMX} we will see that in the former case all prices, apart from the undisturbed one, decay with the same power law whereas the in the latter case the exponent changes with $\lambda$. In fact the first exponent is fully determined by the parameter values {\it after\/} the change. The reason lies in the fact that in this case we have $|\beta^{-}_b|<|\beta^{-}_a|$, and since $|\beta^{-}_a|<|\gamma^{-}_a|$ for any value of $\lambda$, the $\beta^{-}_b$ exponent dominates the extreme behaviour. Note that this feature was subtly present in figure~\ref{figVCM} as well.

We finally show how we can use the result in~(\ref{moment-generating}) to obtain the moments of the exercise time thanks to~(\ref{moments}). In particular, we present the mean value of this first-passage time when the location of the boundary does not depend on $r$, as in the case of binary options, $H_{\ab}=\K$. We will assume that the drift terms are both positive, $\theta_{\ab}>0$, and that the process starts below the barrier. Therefore, we are computing the mean lifetime of a perpetual binary call:  
\begin{eqnarray}
\EE [t^{+}_0-t_0|\FF(t_0)] = 
\frac{1}{\theta_b} \left\{x_0-\frac{\theta_a-\theta_b}{\lambda}\left[1-e^{-\frac{x_0}{\sigma^2_a}\left( \sqrt{\theta^2_a+2\lambda \sigma^2_a}- \theta_a\right)}\right] \right\}.\label{MFPT}
\end{eqnarray}
Note that the result is specially well fitted for the asymptotic analysis of large values of $\lambda$. If we are interested in the case in which the likelihood of a change in the market conditions is very small the following approximate expression becomes more helpful:    
\begin{eqnarray}
\EE [t^{+}_0-t_0|\FF(t_0)]\approx  
\frac{x_0}{\theta_a} \left\{1+ \frac{\lambda}{2}\left(\frac{1}{\theta_b}-\frac{1}{\theta_a}\right)\left[x_0+\frac{\sigma^2_a}{\theta_a}\right] \right\}. \label{MFPTb}
\end{eqnarray}

We must remember that the proper way of performing the computation in~(\ref{moments}) is by assuming that $r$, $\lambda$, $\theta_a$, $\theta_b$, $\sigma_a$ and $\sigma_b$ are free parameters. 
Once we have formally set $r=0$ and obtained~(\ref{MFPT}) we have to recall that this quantity must be evaluated by using the physical not the risk-neutral measure. Therefore we will replace $\theta_{\ab}$ with $\tilde{\theta}_{\ab}$.  

\begin{figure}[tbh] 
{\hfil \includegraphics[width=0.60\textwidth,keepaspectratio=true]{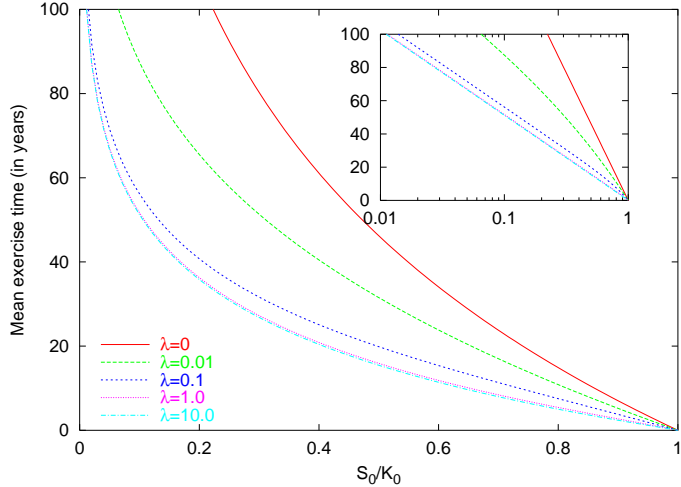}}
\caption{Mean exercise time of a perpetual binary call under drift risk.
We represent the mean lifetime of the option, in terms of the moneyness ($\SI/\K$), for different values of $\lambda$. The values for the parameters are: $\tilde{\theta}_a=1.5\%$, $\tilde{\theta}_b=3\%$ and $\sigma_a=10\%$.}
\label{figMFPT}
\end{figure}

In figure~\ref{figMFPT} we have depicted the mean lifetime of binary calls under the following market conditions:  $\tilde{\theta}_a=1.5\%$, $\tilde{\theta}_b=3\%$ and $\sigma_a=10\%$.
Obviously the fact that the drift is bigger after the transition reduces the mean time as the likelihood of the change increases.

From~(\ref{MFPT}) it becomes evident that the drift of the process is determinant in computing the mean first-passage time, whereas the volatility plays a marginal role: in fact the value of $\sigma_b$ appears only through $\tilde{\theta}_b$. 
This outcome is no longer true if we focus our interest on the second moment of the first-passage time. For the sake of simplicity we will assume that $\tilde{\theta}_a=\tilde{\theta}_b=\tilde{\theta}$, but with $\sigma_a\neq\sigma_b$. We can observe how, even under our previous assumptions, the second moment depends explicitly on the value of $\sigma_a$ and $\sigma_b$: 
\begin{eqnarray*}
\EE [(t^{+}_0-t_0)^2|\FF(t_0)] = \\
\frac{1}{\tilde{\theta}^2}\left\{x_0\left[x_0+\frac{\sigma^2_b}{\tilde{\theta}}\right]-\frac{\sigma^2_b-\sigma^2_a}{\lambda}\left[1-e^{-\frac{x_0}{\sigma^2_a}\left( \sqrt{\tilde{\theta}^2+2\lambda \sigma^2_a}- \tilde{\theta}\right)}\right] \right\},\label{VFPT}
\end{eqnarray*}
an expression that reduces to
\begin{eqnarray}
\EE [(t^{+}_0-t_0)^2|\FF(t_0)]\approx  
\frac{x_0}{\tilde{\theta}^2} \left[x_0+\frac{\sigma^2_a}{\tilde{\theta}}\right]\left\{1+ \frac{\lambda}{2}\left(\frac{\sigma^2_b-\sigma^2_a}{\tilde{\theta}^2}\right) \right\}, \label{VFPTb}
\end{eqnarray}
if we are concerned about small values of $\lambda$. In fact, if we confront~(\ref{MFPTb}) and~(\ref{VFPTb}) we realize that the first correction to the mean first-passage time is governed by the undisturbed second moment of the process. 
\begin{figure}[tbh] 
{\hfil\includegraphics[width=0.60\textwidth,keepaspectratio=true]{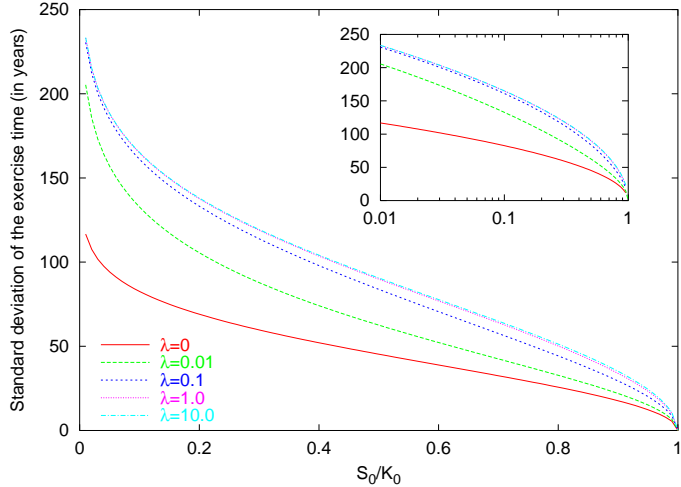}}
\caption{Standard deviation of the exercise time of a perpetual binary call without drift risk.
We represent the variance of the lifetime of the option, in terms of the moneyness ($\SI/\K$), for different values of $\lambda$. The values of the parameters were adjusted in order to keep the drift unchanged: $\tilde{\theta}_a=\tilde{\theta}_b=1.5\%$.}
\label{figVFPT}
\end{figure}

In figure~\ref{figVFPT} we plot an example of the standard deviation of the exercise time which will illustrate the constant drift instance. Clearly this is only feasible if the value of the parameters are well tuned: $\mu_a=2\%$, $\mu_b=3.5\%$, $\sigma_a=10\%$ and $\sigma_b=20\%$. The outcome fits our anticipation, since the possibility of a larger volatility increases the uncertainty about the mean exercise time. 

\section{Conclusions and future work}
\label{Last}
In this article we have considered the implications that the presence in the market of volatility and dividend risk has for option prices. The proposed market model allows random changes in both the dividend-payment rate and the volatility level of stock shares, but in a very specific way: only one change is feasible, and the final market properties are foreseeable. The model, however, is rich and realistic enough to obtain valid financial results. We have focused our attention on the problem of pricing perpetual American options: derivatives with no expiration limit that may be exercised at any time. The absence of maturity in a derivative departs from actual market conditions, but it is a well-accepted approximation used with the purpose of casting light on the way of solving the whole problem. From a physical point of view, the analysis of perpetual options may have even bigger interest than real options because, usually, physical systems do not disappear after a fixed time lag.

Within this framework, we have obtained explicit solutions for pricing derivatives with the most typical pay-offs: vanilla puts and calls, as well as binary options. 
Nevertheless, the applications of our development are not restricted to finance since, as we have pointed out, binary option prices are nothing but classical moment-generating functions of a first-passage time. We have shown the right way of handling these expressions within our set-up. 
We have illustrated our results with a set of practical examples covering the major issues we have analysed. 

The most intriguing aspect of our outcomes is the possible existence, depending on the actual market conditions, of a threshold that limits the probability of change in order to have a valid price. We have found no sound clue that may explain the financial meaning of this behaviour, beyond the fact that in such a case the smooth pasting condition cannot be satisfied. We believe that this feature deserves deeper analysis in a forthcoming work. We aim to extend these results to American derivatives with maturity date as well. It is probable that the solution to this problem will imply the use of approximate procedures and numerical techniques to some extent.

\ack The author acknowledge partial support from the Spanish {\it Ministerio de Ciencia y Tecnolog\'{\i}a\/} under contracts No. BFM2003-04574 and No. FIS2006-05204-E, and by the {\it Generalitat de Catalunya\/} under contract No. 2005 SGR-00515.

\appendix

\section{The financial approach}
\label{SA}
In this appendix we will present a supplementary approach to the issue. Both methods lead to the same results but differ in the way in which they are obtained. This fact can made a technique more  suitable than the other when discussing some properties of the model, like the question of the completeness of the market. This alternative approach is based on the idea that the fair price of an option must be equal to the value of some portfolio made of different securities that mimics the behaviour of the derivative and hedges all the risk. In fact, the market will be complete if we can construct such {\it hedging portfolio\/} for every traded asset of the market. 

The first security to be included in the portfolio is the underlying asset, which will reproduce changes in the option price due to the evolution of the stock price $S$. The second security in our portfolio is a zero-coupon bond, a free-risk monetary asset, with a market price $B$ that satisfies the following differential equation: 
\begin{equation}
dB=r B dt. \label{dB}
\end{equation}
A long position in this security will provide a secure resort where to keep the benefits of an effective investment strategy, whereas a short position in these bonds will allow us to borrow money when we need it. 
These two securities cannot counterbalance all the stochastic behaviour of the price of the option however: not all the influence of $\delta(t)$ and $\sigma(t)$ on the option price may be explained through $S$; and $B$ is fully deterministic. Therefore, we need another security that can account for this contribution to the global risk. Nevertheless, in the most of the cases, markets do not trade such assets, a fact that impel us to consider the inclusion of a {\it secondary\/} option in the portfolio: a derivative of the same nature of $P(t,S;\KK)$, but with a different set of contract specifications, $Q(t,S;\KK')$. In particular we will focus on the case in which they share the same expiration date but differ on the striking price. 
Note that we will assume as well that the option price is a function of the {\it current\/} value of the underlying. 

Let us write down $P$ as a mixture of $\nu$ shares $S$, $\phi$ units of the riskless security $B$, and $\psi$ secondary options $Q$:
\begin{equation}
P=\nu S + \phi B + \psi Q.
\label{Pdef}
\end{equation}
The variation in the value of the portfolio, due to the market evolution of its components and the received dividends, fulfils the following relationship:
\begin{equation}
dP=\nu dS + \nu \delta S dt + \phi dB + \psi dQ, \label{dP}
\end{equation}
where we have taken into account that $\nu$, $\phi$, $\psi$ and $\delta$ are predictable processes, and that we adopt a {\it self-financing strategy\/}, in which there is no net cash flow entering or leaving the replicating portfolio. This differential change must equal the expression obtained after applying the rules of It\^o calculus on the price of the option:     
\begin{equation}
dP=\partial_t P dt+\partial_{S} P dS+\frac{1}{2} \sigma^2 S^2 \partial_{SS}^2 P dt+ \Delta P d{\mathbf 1}_{t\geq \tau}, \label{ItodP}
\end{equation}
where
\begin{equation*}
\Delta P \equiv P_b(t,S;\KK)-P(t,S;\KK),
\end{equation*}
and $P_b(t,S;\KK)$ is the price of the option after the jump. The differential of an indicator with a random variable in its argument may seem a bizarre object. However, it is mathematically well defined. In fact, ${\mathbf 1}_{t\geq \tau}$ is a submartingale under our measure and, by virtue of the Doob-Meyer decomposition theorem,  $d{\mathbf 1}_{t\geq \tau}$ can be expressed as a sum of two terms, $d{\mathbf 1}_{t\geq \tau}=dA + dM$, an increasing adapted process $A$, 
\begin{equation}
A(t)=\lambda \min(\tau-t_0,t-t_0), 
\label{Adef}
\end{equation}
and a cadlag martingale $M$,
\begin{equation}
M(t)=1-\lambda \EE[\tau-t_0|\FF(t)]. 
\label{Mdef}
\end{equation}
The key point is that $d{\mathbf 1}_{t\geq \tau}$ is a stochastic magnitude, independent of $dW$, which does not directly contribute to the variation of the stock price $dS$. Therefore, it is a source of risk that cannot be explained in terms of the random evolution of the underlying asset.

The combination of~(\ref{dP}) and~(\ref{ItodP}), together with~(\ref{dB}), lead to: 
\begin{eqnarray}
\partial_t P dt+\partial_{S} P dS+\frac{1}{2} \sigma^2 S^2 \partial_{SS}^2 P dt+ \Delta P d{\mathbf 1}_{t\geq \tau}= \nonumber \\
\nu dS+  \delta S dt + r \phi B dt + \psi dQ. \label{PDE_1}
\end{eqnarray}
Now, we can proceed with $dQ$ in an analogous way,
\begin{equation}
dQ=\partial_t Q dt+\partial_{S} Q dS+\frac{1}{2} \sigma^2 S^2 \partial_{SS}^2 Q dt+ \Delta Q d{\mathbf 1}_{t\geq \tau}, \label{dQ}
\end{equation}
where the natural definition of $\Delta Q$,
\begin{equation*}
\Delta Q=Q_b(t,S;\KK')-Q(t,S;\KK'),
\end{equation*}
has been used. In order to recover a deterministic partial differential equation we must guarantee that terms containing the stochastic magnitudes $dS$ and $d{\mathbf 1}_{t\geq \tau}$ mutually cancel out. Therefore we must demand that
\begin{equation*}
\nu=\partial_{S} P - \psi \partial_{S} Q,
\end{equation*}
a condition named {\it delta hedging\/}, and also that
\begin{equation*}
\psi=\frac{\Delta P}{\Delta Q},
\end{equation*}
which is usually referred as {\it vega hedging\/}. The previous hedging conditions reduce~(\ref{PDE_1}) to
\begin{eqnarray}
\partial_t P +\frac{1}{2} \sigma^2 S^2 \partial_{SS}^2 P -\delta S \partial_{S} P = \nonumber \\
  r \phi B  + \frac{\Delta P}{\Delta Q} \left( \partial_t Q +\frac{1}{2} \sigma^2 S^2 \partial_{SS}^2 Q -\delta S \partial_{S} Q \right), \label{CBnD}
\end{eqnarray}
an expression that still involves $B$, which is not an inner variable of option prices $P$ and $Q$ in our set-up.\footnote{In fact, some authors replaces $t$ with $B$ as the free variable.} This problem can be fixed using the definition of the portfolio in~(\ref{Pdef}) and the {\it vega hedging\/} together,
\begin{equation*}
\phi B=P-\nu S - \psi Q=P -\left(\partial_{S} P - \frac{\Delta P}{\Delta Q}\partial_{S} Q \right)S - \frac{\Delta P}{\Delta Q} Q.
\end{equation*}
The replacement of $\phi B$ in~(\ref{CBnD}) leads to
\begin{eqnarray*}
\partial_t P +\frac{1}{2} \sigma^2 S^2 \partial_{SS}^2 P -rP +(r-\delta) S\partial_S P
=\\
\frac{\Delta P}{\Delta Q} \left( \partial_t Q+\frac{1}{2} \sigma^2 S^2 \partial_{SS}^2 Q-rQ +(r-\delta)S\partial_S Q\right).
\end{eqnarray*}
This formula implies the existence of an arbitrary function $\chi$, which uncouples the problem of finding $P$ and $Q$:
\begin{equation}
\chi=\frac{1}{\Delta P}\left(
\partial_t P +\frac{1}{2} \sigma^2 S^2 \partial_{SS}^2 P -rP +(r-\delta) S\partial_S P \right). \label{chi}
\end{equation}
Obviously the same formula is valid for the secondary option, merely by replacing $P$ with $Q$. This proves that the option $Q$ removes the remaining risk and therefore completes the market.

The financial interpretation of $\chi$ is discussed in more depth in~\cite{H98} and~\cite{MM04}. The must known facts are four: (i) $\chi$ depends on how every investor measures the volatility and dividend risk, what prevents us from fixing it, (ii) prior to the change it must be negative defined or otherwise the market will show arbitrage opportunities, (iii) after the change must be zero and, finally, (iv) the choice $\chi=-\lambda{\mathbf 1}_{t < \tau}$ avoids the so-called statistical arbitrage, the growth of the expected value of the {\it discounted\/} price of the option, $\hat{P}=e^{-r(t-t_0)}P$. The violation of this condition is against the capital asset pricing model which states that any hedged portfolio with a zero market risk must have an expected return equal to the risk-free rate.
It can be easily shown that this choice for $\chi$ lead to the desired property: from~(\ref{ItodP}) and~(\ref{chi}) it follows that $d\hat{P}$ can be decomposed in two parts,
\begin{equation*}
d\hat{P}= \partial_{S} P dF+ \Delta P dG, 
\end{equation*}
where $F$ was defined in~(\ref{defF}) and $dG=\chi dt+d{\mathbf 1}_{t\geq \tau}$. On the one hand $F$ is (and must be) a martingale under our measure ---see~(\ref{martF}). On the other hand, if we set 
\begin{equation}
\chi dt=-dA=-\lambda{\mathbf 1}_{t < \tau}dt,
\label{Thechi}
\end{equation}
with $A$ defined as in~(\ref{Adef}), we will have that $G$ equals $M$ ---see~(\ref{Mdef})---, and therefore becomes a martingale as well, what proves our previous statement. Note however that in general $\chi$ may depend on the two independent variables $t$ and $S$, on market properties $r$, $\delta$, $\sigma$ and $\tau$, but also on the common contract specification $T$. 

In fact, financial arguments ---see for instance \cite{M73}--- lead to the conclusion that the natural time variable in option pricing problems is not $t$ but $\bar{t}=T-t$, the time to maturity. This convention, for instance, turns the pay-off function into an initial condition in European-like contract problems $P(\bar{t}=0,S=S(T);K)=X(S(T);K)$ and removes explicit temporal dependence from perpetual derivative problems, also in the optimal exercise barrier of American options $\lim_{\bar{t}\rightarrow \infty} H(\bar{t})=H_{\ab}$. This reduces~(\ref{chi}) to an ordinary differential equation:
\begin{equation}
\frac{1}{2} \sigma^2 S^2 \frac{d^2 P}{d S^2} +(r-\delta) S\frac{d P}{d S} -(r-\chi) P=\chi P_b. \label{odeP}
\end{equation}
The solution if the jump has taken place does not depend on $\chi$ since the equation to be solved is~\cite{BAW87}:
\begin{equation}
\frac{1}{2} \sigma^2_b S^2 \frac{d^2 P_b}{d S^2} +(r-\delta_b) S\frac{d P_b}{d S} -r P_b=0. 
\label{odePb}
\end{equation}
Here $\chi$ becomes meaningless and thus we can freely set $\chi=0$. The general solution of~(\ref{odePb}) for $\delta_b \neq 0$ reads
\begin{equation*}
P_b=C_{1} S^{\beta^{+}_b}+C_0 S^{\beta^{-}_b},
\end {equation*}
where $\beta^{\pm}_b$ were defined in the main text ---cf~(\ref{betapm})---, and constants $C_1$ and $C_0$ can be obtained by demanding the solution to fulfil boundary conditions~(\ref{continuity}) and~(\ref{smooth}), if applicable: e.g. $C_1=0$ and $C_0=1/\K^{\beta^{-}_b}$ lead to a perpetual American binary put, $C_1=(H_b-K)/H_b^{\beta^{+}_b}$ and $C_0=0$ to a perpetual American vanilla call, and so forth. 
The solution if the jump has not taken place yet {\it does\/} depend on the arbitrary function $\chi$. The development in the main text is consistent with~(\ref{Thechi}), i.e. $\chi = -\lambda$. In this case~(\ref{odeP}) reduces to:  
\begin{equation}
\frac{1}{2} \sigma^2_a S^2 \frac{d^2 P}{d S^2} +(r-\delta_a) S\frac{d P}{d S} -(r+\lambda) P=-\lambda P_b, 
\label{odePa}
\end{equation}
whose general solution is
\begin{equation}
P=C_{3} S^{\gamma^{+}_a}+C_2 S^{\gamma^{-}_a} +\lambda\left[\frac{C_{1}}{\lambda+\ell^{+}}S^{\beta^{+}_b} +\frac{C_{0}}{\lambda+\ell^{-}} S^{\beta^{-}_b}\right],
\label{odePaSol}
\end {equation}
whenever $\gamma^{+}_a\neq\gamma^{-}_a$ and $\gamma^{\pm}_a\neq\beta^{\pm}_b$. Expressions for $\gamma^{\pm}_a$ and $\ell^{\pm}$ have been already introduced in the main text, in~(\ref{gamma}) and~(\ref{ell}) respectively. Once again the arbitrary constants in~(\ref{odePaSol}) must be evaluated on the basis of contract specifications: for example 
$C_3=\ell^{+}/(\lambda+\ell^{+})\K^{\gamma^{+}_a}$, $C_1=1/\K^{\beta^{+}_b}$ and $C_2=C_0=0$ reproduce the price of a perpetual American binary call.

Clearly, the previous method is well suited to discuss most of the financial assumptions and implications of the market model but dilutes the connections of the issue with other branches of knowledge. The fact that the solution of~(\ref{odeP}) is related to a first-passage time problem may be not so evident at first glance, for instance. The way of recovering the results corresponding to the use of the physical measure is also a delicate question. These and other reasons impelled us to follow in the main text a development which is akin to physicists, and to leave the mathematical finance approach to this appendix.

\section{The smooth pasting condition}
\label{SB}
In this appendix we analyse the properties of the smooth pasting condition for vanilla options: formula~(\ref{HH}) in the main text. Let us first define the auxiliary function $\GG^{\pm}(\eta)$, 
\begin{equation*}
\GG^{\pm}(\eta) = A^{\pm} \eta^{1-\beta^{\pm}_b} \mp B^{\pm} -\eta^{-\beta^{\pm}_b}, 
\end{equation*}
with
\begin{eqnarray*}
A^{\pm}&=& \frac{\beta^{\pm}_b}{\beta^{\pm}_b-1} \frac{\gamma^{\pm}_a-1}{\gamma^{\pm}_a}, \\
B^{\pm}&=& \frac{\pm 1}{\beta^{\pm}_b-1}\frac{\lambda}{\lambda+r+ \sigma^2_a \gamma^{\pm}_a\beta^{\pm}_b/2}. 
\end{eqnarray*}
The study of this function is relevant in our framework since $H_a$ fulfils $\GG^{\pm}(H_a/H_b)=0$ ---remember that $H_b$ was introduced in~(\ref{Hb}). The properties of $\beta^{\pm}_b$ and $\gamma^{\pm}_a$ impel us to treat separately call and put optimal boundaries. For call options we will have that $\beta^{+}_{\ab}\geq 1$ and $\gamma^{+}_a\geq 1$. We will dismiss the particular case $\beta^{+}_b=1$ because it was analysed in~\Sref{S4}. 
Moreover, the only scenario in which $\gamma^{+}_a= 1$ corresponds to a very specific situation: $\beta^{+}_a=1$ and $\lambda=0$, the jump never takes place. Therefore, $A^{+}$ and $B^{+}$ are bounded and positive constants, $\GG^{+}(\eta)$ fulfils $\lim_{\eta \rightarrow 0}\GG^{+}(\eta)=-\infty$ and  $\lim_{\eta \rightarrow \infty} \GG^{+}(\eta)=-B^{+}$, and it shows a single maximum at $\eta=\eta_M$:
\begin{equation*}
\eta_M=\frac{\gamma^{+}_a}{\gamma^{+}_a-1}. 
\end{equation*}
That means that the equation $\GG^{+}(\eta)=0$ may have two solutions, one solution or no solution at all, depending on the value of $\GG^{+}(\eta_M)$. When the parameters lead to
\begin{equation}
\left(\frac{\gamma^{+}_a-1}{\gamma^{+}_a}\right)^{\beta^{+}_b} >\frac{\lambda}{\lambda+r+ \sigma^2_a \gamma^{+}_a\beta^{+}_b/2}, 
\label{Cp}
\end{equation}
for any value of $\lambda$ we will have two formal solutions, but we can easily reject one of them by analysing the properties of the boundary for large and small values of $\lambda$. In fact, condition~(\ref{Cp})  is always satisfied for small enough values of $\lambda$ since  
\begin{equation*}
\left(\frac{\beta^{+}_a-1}{\beta^{+}_a}\right)^{\beta^{+}_b} >0.
\end{equation*}
The previous statement is false if $\beta^{+}_a = 1$, i.e. when the stock pays no dividend before the change. In this case we will have found no parameter choice which fulfils~(\ref{Cp}) whenever $\beta^{+}_b \neq 1$. In particular, it is easy (but tedious) to show that if $r>0$ and $\delta_a=0$ then
\begin{equation*}
\left(\frac{\gamma^{+}_a-1}{\gamma^{+}_a}\right)^{2} <\frac{\lambda}{\lambda+r+ \sigma^2_a \gamma^{+}_a}, 
\end{equation*}
for all $\lambda>0$. 

Let us assume again that $\beta^{+}_{\ab} > 1$. Under these assumptions, and depending on the market value of the parameters, may exist a threshold for $\lambda$, $\bar{\lambda}$, above which no optimal boundary can be defined:
\begin{equation*}
\left(\frac{\bar{\gamma}^{+}_a-1}{\bar{\gamma}^{+}_a}\right)^{\beta^{+}_b} =\frac{\bar{\lambda}}{\bar{\lambda}+r+ \sigma^2_a \bar{\gamma}^{+}_a\beta^{+}_b/2}, 
\end{equation*}
In fact, for $\lambda=\bar{\lambda}$, we will have a single solution:
\begin{equation*}
H_a=\frac{\bar{\gamma}^{+}_a}{\bar{\gamma}^{+}_a-1} H_b. 
\end{equation*}

This situation is also present in the estimation of the optimal boundary for puts. Here we have $\beta^{-}_{\ab}<0$ and $\gamma^{-}_a<0$, what leads to the conclusion that $A^{-}$ and $B^{-}$ are also bounded and positive constants, that the auxiliary function $\GG^{-}(\eta)$ fulfils $\lim_{\eta \rightarrow 0} \GG^{-}(\eta)=B^{-}$, $\lim_{\eta \rightarrow \infty} \GG^{-}(\eta)=A^{-}$, and that $\eta_m$,
\begin{equation*}
\eta_m=\frac{|\gamma^{-}_a|}{|\gamma^{-}_a|+1}, 
\end{equation*}
is a global minimum of $\GG^{-}(\eta)$. The condition for the existence of a solution for the optimal boundary problem now reads:
\begin{equation}
\left(\frac{|\gamma^{-}_a|+1}{|\gamma^{-}_a|}\right)^{|\beta^{-}_b|} <\frac{\lambda+r+ \sigma^2_a |\gamma^{-}_a||\beta^{-}_b|/2}{\lambda}, 
\label{Cm}
\end{equation}
and, like in the call case, it will be satisfied for small values of $\lambda$ because 
\begin{equation*}
\left(\frac{|\beta^{-}_a|+1}{|\beta^{-}_a|}\right)^{|\beta^{-}_b|} < \infty.
\end{equation*}
Therefore we may have a threshold in $\lambda$ depending on the market properties as well. Let us show this behaviour through a practical example. Consider the case $\beta^{+}_b=2$. Here the constraint~(\ref{Cm}) is equivalent to demanding $(r+\theta_a)|\gamma^{-}_a|+r>0$. This means that for $\theta_a \geq -r$ we will always find a optimal boundary, whereas for $\theta_a < -r$ there will be solution only in the range
\begin{equation*}
\lambda\leq\bar{\lambda}= \frac{r^2}{(|\theta_a|-r)^2}\left(|\theta_a|-r + \frac{\sigma^2_a}{2}\right).
\end{equation*}
The value of $H_a$, when it exits, follows:
\begin{equation*}
H_a=\frac{K}{2} \frac{|\gamma^{-}_a|}{|\gamma^{-}_a|+1} \left(1+\sqrt{\frac{r+(r+\theta_a)|\gamma^{-}_a|}{\lambda +r  +(\lambda +r +\theta_a)|\gamma^{-}_a|}}\right).
\end{equation*}
Note that $H_a$ shows the right limiting properties $\lim_{\lambda \rightarrow 0} H_a=K |\beta^{-}_a|/(|\beta^{-}_a|+1)$ and $\lim_{\lambda \rightarrow \infty} H_a=H_b=K/2$.

We have a plausible explanation of the absence of solution for any value of $\lambda$ in the case of calls on  non-paying dividend shares: the perpetual American call behaves like the underlying and thus there is no optimal boundary. The financial interpretation of the existence of threshold $\bar{\lambda}$, but only under some market conditions, is elusive.

\end{document}